# *Einstein's Happiest Moment: The Equivalence Principle*

Paul Worden
Stanford University
pworden@stanford.edu

and

James Overduin
Towson University and Johns Hopkins University
joverduin@towson.edu

Einstein's happiest thought was his leap from the observation that a falling person feels no gravity to the realization that gravity might be *equivalent* to acceleration. It affects all bodies in the same way because it is a property of spacetime — its curvature — not a force propagating through spacetime (like electromagnetic or nuclear forces). When expressed in a way that is manifestly independent of the choice of coordinates, this idea became General Relativity. But the ground for what is now known as the "equivalence principle" was laid long before Einstein, affording a fascinating example of the growth of a scientific idea through the continuous interplay between theory and experiment. That story continues through the present era, with improved techniques and measurements, and in the new environment of space. Theoretically, equivalence is now understood to rank with Lorentz invariance as one of nature's most fundamental principles. Violations of equivalence are generically predicted by attempts to unify gravity with "non-geometrical" fundamental interactions, but there is no consensus on the form that these violations will take. Experimental tests of equivalence thus face the challenge of maximizing both sensitivity and diversity of test materials.

## **The Equivalence Principle**

In physics a *Principle* is something we believe but do not yet understand. Some Principles, for example the Cosmological Principle, originate primarily in philosophy, but the Equivalence Principle (EP) began with an experimental fact. That two conceptually independent quantities — inertial mass (the "$m$" in $F = ma$) and gravitational mass (the "$m$" in $F = GMm/r^2$) — should be identical is arguably one of the strangest facts in physics. No such identity holds for any of the other fundamental forces. But this experimental fact, elevated to a postulate, permits gravity to be interpreted as geometry. If it were in any degree not true, different geometries would be needed for different material objects, violating the Principle of Parsimony (Occam's Razor) — another philosophical concept.

In contrast a *Theory* is a tool for understanding a phenomenon. It is a well-defined framework for testing and refining ideas. We compare the consequences of the theory with observations, and if they agree, then *maybe* the theory is true, but at least it describes whatever is really happening. Otherwise, we may need to fix the theory or get a new one. An *Experiment* is a formal observation. These ideas and the scientific method have been well described elsewhere.

Historically, theories tend to evolve to give better and better descriptions of their subject, because it is easier to patch them up than make new ones. They may become accurate enough to be practically useful — even for designing new experiments, which may lead to completely different theories. Theories continue to evolve until suddenly they don't and a new, better theory appears. This was the case with Newton's *Principia* in 1687, and again with Einstein's General Relativity (GR) in 1915.

**Milestones**

The earliest hints of something like equivalence came from Aristotle. In classical philosophy, figures such as Socrates, Plato, and Aristotle were teachers who promoted their own ideas for a living, and were regarded as final authorities by their followers. The cultural environment was essentially conservative, but provided them opportunity for speculation and questioning ... as long as it wasn't too serious. Discussing the nature of things was less controversial than politics or religion and everyone had experience of it; anyone could become an expert, since no one was. Greek physics, in the modern sense, was almost entirely theoretical. Philosophy emphasized reason over demonstration, encouraged by successes in geometry. The concepts and tools needed to perform careful experiments did not exist. But observation was not neglected.

A particularly keen observer, Aristotle wrote: "No one can say why something that is moving should stop anywhere; why should it stop here rather than there? Therefore a thing will either be at rest or must move forever, unless something more powerful get in its way."[1] The often-quoted first sentence suggests that he had a concept of inertia — why indeed should something ever stop? The second, some idea of force. But it was all too new, and in this quote Aristotle was not interested in moving bodies at all; instead, he was using his ideas about motion to argue against the existence of a vacuum. Elsewhere he wrote, "We see that, other things being equal, heavy and light bodies move with unequal velocities ... in the ratio which their magnitudes have to each other." This makes little sense for falling bodies in vacuum; but from the context it is clear he is discussing motion in a medium. In such a medium, objects do fall at a terminal velocity roughly proportional to their weight.[2,3,4]

Aristotle was referring to bodies' *speeds*, not their *accelerations*, a concept that did not exist in classical thought. But he had intuitive notions of inertia, weight and resistance in addition to the tools of ratios and geometry. Given his authority and the incompleteness of existing ideas, his successors can be forgiven for misinterpreting him about free fall in a medium. But the misunderstanding has persisted until modern times.[5]

In his philosophical poem *On the Nature of the Universe [De Rerum Natura],* Lucretius (96 – 55 B.C.) wrote: "Through undisturbed vacuum all bodies must travel at equal speed, though impelled by unequal weights."[6] Again this startling statement must be understood in context. It was not a premonition of Newton's laws or the Equivalence Principle, but was clearly inspired by, and a challenge to, Aristotle. What Lucretius was really doing here was preparing for a defense of Epicurean physics, in particular the doctrine of the "swerve" by which complexity emerges from an endlessly falling sea of atoms.[7]

Half a millennium later, perhaps the best-known successor and critic of Aristotle was Johannes Philoponus of Byzantium. Explicitly contradicting Aristotle, Philoponus noted that bodies of different weight do in fact move similarly: "The difference in time is a very small one."[8] Also, "... it is necessary to assume that some incorporeal motive force is imparted by the projector to the projectile [and] ... if one imparts ... forced motion to an arrow or a stone the same degree of motion will be produced much more readily in a void than in a plenum. And there will be no need of any agency external to the projector."[9] A concept resembling inertia and/or momentum is here — Philoponus called it impetus — which is added to the object, and decays through resistance of the medium. There is no clear hint of a property of the object itself resisting motion. Unfortunately, Philoponus's theological ideas also met with resistance, which limited the spread of his philosophy. Cultural inertia prevailed, but the concept of impetus survived, laying the ground for Galileo.

Galileo arguably invented experimental physics, but not every experiment attributed to him actually occurred. Historians have cast doubt on his claims to have compared the periods of otherwise identical pendulums made of cork and lead.[10] The legendary drop test at the Leaning Tower of Pisa using a lead musket ball and an iron cannonball was probably just that — legendary. If it did happen, it was likely an informal lecture demonstration rather than a formal experiment, or it may have been conflated with a similar experiment performed elsewhere.[11] But Galileo carefully developed the law of falling bodies from laboratory experiments with inclined planes, so he knew what the result of a drop test would be without actually doing it. Galileo had a nearly modern idea of inertia — he invented the notion — and a background suited to questioning, mathematics, and experimentation.

Galileo was generously supported by the Medici family and could afford his experiments using simple apparatus. Like Philoponus, Galileo encountered significant resistance to his ideas, but the social context of his era allowed them to become widespread anyway, perhaps driven partly by the notoriety surrounding his trial. He had many contemporaries and colleagues with similar interests. Most of the conceptual prerequisites for the idea of equivalence — and modern physics — were in place by this time, including a better understanding of accelerated motion.[12] It remained to clarify them and put them together.

The EP can be said to have originated with Newton, as did experiments to test it. As he stated in Definition I of the *Principia*: mass is "known by the weight of each body, for it is proportional to the weight, as I have found by experiments on pendulums, very accurately made."[13] Newton was first to use use the terms "force" and "inertia" in their modern meanings: "The *vis insita*, or innate force of matter, is a power of resisting, by which every body, as much as in it lies, continues in its present state, whether it be of rest, or of moving uniformly forwards in a right line ... this may be called inertia (*vis inertiae*) or force of inactivity." He continued with a definition of force, concluding: "And thus everything may be subjected to experiment." Newton recognized that the *composition* of the test bodies might be significant: his pendulum bobs were empty boxes containing equal weights of wood, gold, silver, lead, glass, salt and wheat.[14] Even more brilliantly, he perceived that Nature carries out another kind of "pendulum test" for free. From observations available to him, he was able to infer that the Earth and Moon, as well as Jupiter and its Galilean moons, fall toward the Sun with accelerations that are equivalent to about a part in a thousand,[15] a limit that was later improved by Pierre-Simon Laplace to a few parts in $10^7$.[16]

Friedrich Wilhelm Bessel made the next major contribution by duplicating Newton's pendulum experiments with the explicit goal of reducing the uncertainty in the EP. He also extended the range of test-mass materials, employing pendulum bobs made of gold, silver, lead, iron, zinc, brass, marble, clay, quartz, water, meteoritic iron and even a stony meteorite. As was contemporary practice, Bessel compared the lengths of the seconds pendulum for each material, claiming an accuracy of about one part in 60,000.[17]

Beginning in 1891, Loránd Eötvös pioneered an entirely new and more sensitive test using a torsion gradiometer, an instrument devised for geophysical observations and containing pairs of test masses suspended from a torsion fiber. He noted that the masses are subject not only to the gravitational attraction of the Earth and Sun, but also to the centrifugal force caused by the Earth's rotation as well as its orbital motion around the Sun. Thus, different test bodies should twist the torsion fiber by different amounts if they have different ratios of gravitational to inertial mass. Using brass, glass, antimony, and cork as test materials, he concluded that any such difference is less than about five parts in $10^8$.[18]

Fascinating experiments on what would eventually become the EP continued through the period immediately preceding Einstein's discovery of GR, many inspired by Eötvös but now mostly forgotten. These included efforts to measure the directionality of gravitational attraction on crystals, differences between dissolved and crystalline states, radioactive and non-radioactive materials, and the like.[19,20,21,22]

## **General Relativity**

Notably absent after Galileo's time were meaningful tests of the EP using falling bodies. To detect a difference in acceleration of two objects to one part in $10^9$ by timing, one needs to measure their time of fall to one part in $10^9$. From the Leaning Tower of Pisa, this would require a few billionths of a second to match Eötvös — never mind air resistance. Timing techniques accurate enough to compete with Newton's, Bessel's or Eötvös' experiments did not exist before the late 20th century. Even today, it would be difficult to match Eötvös' sensitivity with a free-fall experiment, owing to the short time of fall and difficulties in releasing the masses simultaneously, measuring them, and protecting them from disturbances.

Falling test masses were however central to a famous *thought* experiment conducted by Einstein at his desk in the Swiss patent office in 1907:[23]

> When I was busy (in 1907) writing a summary of my work on the theory of special relativity ... I also had to try to modify the Newtonian theory of gravitation such as to fit its laws into the theory ... At that moment I got the happiest thought of my life ... Because *for an observer in free-fall from the roof of a house there is during the fall — at least in his immediate vicinity — no gravitational field*. That is, if the observer lets go of any bodies, they remain relative to him, in a state of rest or uniform motion, independent of their particular chemical or physical nature. [italics added]

Einstein's happiest moment is illustrated in Fig. 1. Gravitation is (locally) equivalent to acceleration. The qualification "local" is important here, because if the observer lets go of two bodies that are far enough apart, and fall for long enough, she will see those bodies drift together as their geodesic paths converge on the center of the Earth. This lateral drift proves that the observer is in a gravitational field, and not merely accelerating. So the equivalence between gravitation and acceleration is not global.

Dazzling as it was, Einstein's insight that a freely falling observer feels no gravity can be seen as a logical development of Galileo's equally bold realization 300 years earlier that a uniformly moving body feels no force. Combined with the Lorentz invariance of electromagnetism, Galileo's insight paved the way for Special Relativity (SR). With the requirement of invariance under general coordinate transformations (General Covariance), Einstein's EP led him to General Relativity. Its appeal lies not only in its simplifying power (a force reduced to geometry) but in the way it explains why all bodies behave the same way in the same gravitational field (the universality of free fall). Newton had recognized the importance of this question but swept it under the rug by assuming that an object's gravitational mass must, for some mysterious reason, be identical to the inertial mass "$m$" in the law of inertia $F = ma$ that governs all forces. The EP removes this mystery. Gravitation can be transformed away by accelerating. Therefore it must be independent of the bodies themselves, and be a property of the spacetime through which they move. The relevant property is *curvature*. It is especially warping (or "dilation") in the time direction that we feel as gravity, because its effects must be multiplied by the speed of light $c$ to give them the same units as space curvature. (No other speed will do, because no other speed is constant for all observers!)

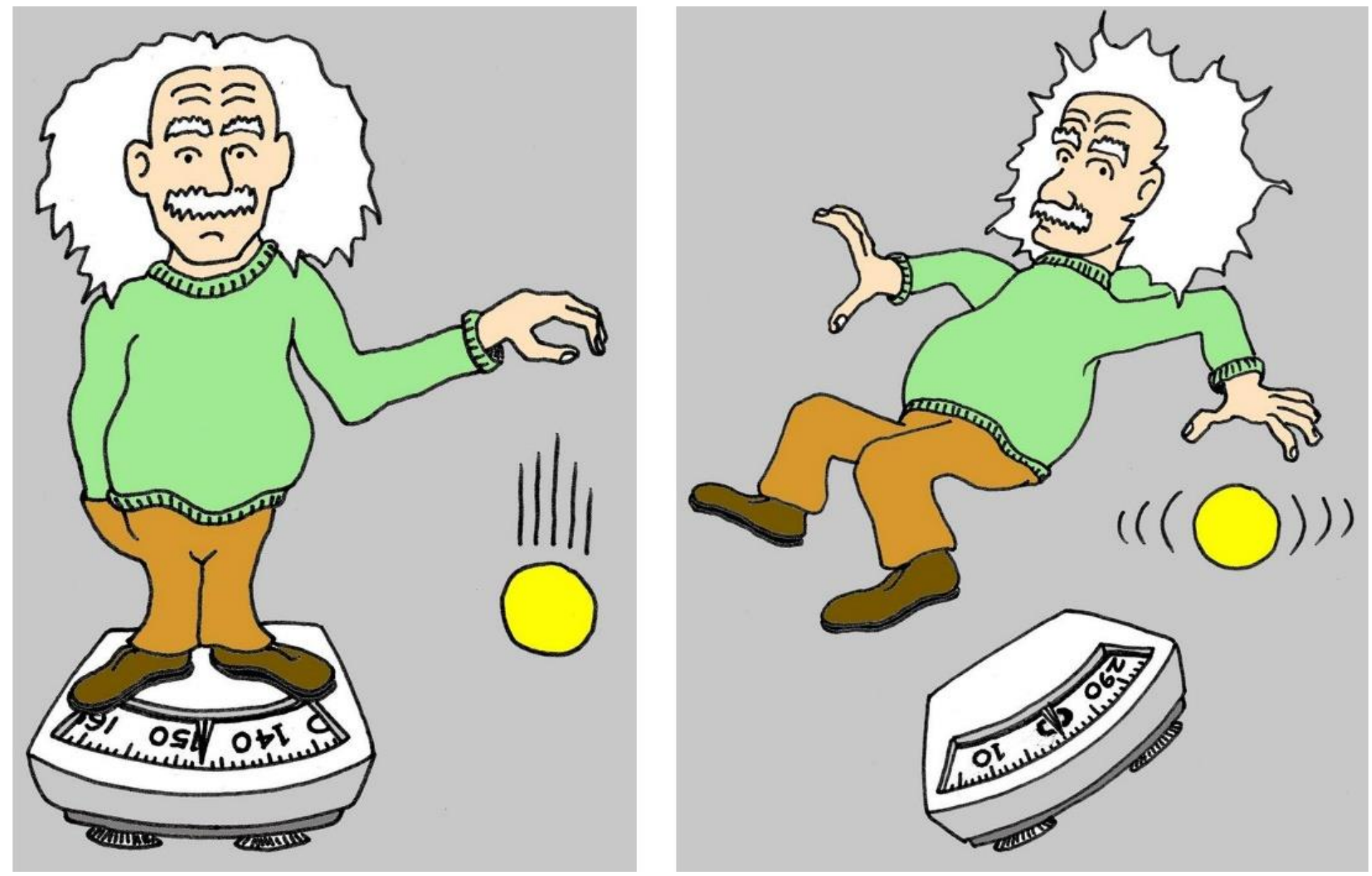

Fig. 1: Einstein's happiest thought: for a freely falling observer, there is no gravity

The statement that gravitational and inertial mass are equal, or equivalently the universality of free fall, is often referred to as the "weak" EP (WEP). It motivated Einstein, but it is important to recognize that he was saying more than this. According to the WEP, there is no way to distinguish (again locally) between falling in a gravitational field $g$, or accelerating upwards with acceleration $g$. From this, Einstein inferred that the same equivalence holds locally for *all non-gravitational phenomena,* not just falling. This is sometimes referred to as Einstein's EP (EEP). Protons and electrons obey the WEP when they fall with the same acceleration. But when an *atom* composed of a proton and electron falls with that same acceleration, it is obeying the EEP, because it also contains electromagnetic binding energy (to say nothing of quarks and the strong force). In the 1960s, Schiff conjectured that any complete and self-consistent theory that satisfies the WEP must also satisfy the EEP,[24] which was shown to be true for a large class of theories by Lightman and Lee in 1972.[25] But a general proof has not been given, and counter-examples exist, suggesting that the validity of Schiff's conjecture probably depends on the meaning of the phrase "complete and self-consistent." Will has shown that, in practice, the EEP boils down to the combination of the WEP with local Lorentz invariance and local position invariance.[26] Theories that satisfy the EEP can be shown to be necessarily "geometric" or "metric theories" of gravity, in which freely falling bodies follow geodesics of a single, symmetric spacetime metric and all of physics (possibly excepting gravity itself) reduces to SR in freely falling reference frames.

There is a third version of the EP, known as the "strong" EP (SEP), which goes further and extends the same equivalence to all *gravitational*, as well as non-gravitational phenomena. When the Earth and Moon fall toward the Sun with the same acceleration, they are satisfying not only the WEP and EEP, but also the SEP, because the Earth contains more gravitational binding energy than the Moon. The SEP is not necessary for the derivation of GR; the EEP is sufficient. GR is however consistent with the SEP, and is in fact currently the only successful theory of gravity which satisfies the SEP. Theories that violate the SEP generally contain new fields which couple to the gravitational field in different ways; examples include the scalar field in Brans-Dicke theory, some cosomological models of dark energy or "quintessence," and "dilaton fields" in most versions of string theory.[27] The profusion of nomenclature and acronyms is unfortunate, as it obscures the fact that equivalence holds a place at the foundation of modern physics that is comparable in importance to Lorentz invariance. It separates "geometrizable" interactions (gravity) from those that are not (particle physics), posing a severe challenge to all would-be unified theories of fundamental interactions.

The classical tests of GR proposed by Einstein have some bearing on the EP, but it should be kept in mind that the EP is a *postulate*, not a prediction of GR. Thus tests of the EP are, in a strong sense, more fundamental than tests of GR. It is possible for a theory to satisfy the EP and still disagree with GR, but GR cannot violate the EP. Einstein's first such test, gravitational redshift, is now understood as a test of the EEP, not GR, and specifically of the requirement of local position invariance. Clocks at different locations in a gravitational field are effectively accelerating with respect to one another, producing a Doppler-like shift which has been detected in numerous experiments, most notably by the Gravity Probe-A experiment in 1976, which verified it to a precision of 7 parts in $10^5$.[28] Modern GPS receivers must compensate for this effect when combining signals from orbiting satellites with receivers on Earth.

Einstein's two other classical tests, light deflection and perihelion shift, are now routinely analyzed using a theoretical framework known as the Parametrized Post-Newtonian (PPN) formalism that was introduced by Nordtvedt in 1968.[29] It uses up to ten new phenomenological parameters to allow for theories of gravity that depart from GR but are still metric in form. Experiments constrain these parameters, like $\gamma$ (related to space curvature) and $\beta$ (related to nonlinearity).[30] Current constraints are of order parts in $10^4$. The phenomenon of light deflection (or gravitational lensing) does serve as a test of the EP as well as GR, since a beam of light passing horizontally over the floor will approach the floor if the room is accelerating upwards. Therefore, according to the EEP, the same thing must occur in a gravitational field. However, the EEP explains only half of the total deflection angle; the rest is due to space curvature. Perihelion precession is more purely a test of GR.

An even more ambitious phenomenological framework has been introduced in the twenty-first century by Kostelecký and collaborators.[31] Known as the Standard-Model Extension (SME), it incorporates not only GR but also the entire standard model of particle physics (SM), and contains hundreds of free parameters allowing for violations of Lorentz invariance as well as the EP. Like the PPN formalism, the SME has prompted much new work on the EP by experimentalists and theorists alike.[32,33]

## **Modern Experiments**

The publication of Einstein's theory of GR in 1918 stimulated experimentalists to improve the sensitivity of existing techniques. Zeeman described experiments with precision balances and a torsion pendulum.[34] Eötvös, Pekár and Fekete published a detailed account of experiments with Eötvös' apparatus, claiming a sensitivity of a few parts in $10^9$ with test samples including water, copper, platinum, copper sulfate, magnesium-aluminum alloy, wood and tallow.[35] Potter performed an improved version of Bessel's experiment, reaching parts per million accuracy.[36] But instrumentation had not improved much over what was available in the previous century. Real improvement had to wait for the electronic sensors and automated systems that followed the end of WWII and the dawn of the space age. References to the term "equivalence principle" in the literature increased by more than an order of magnitude from the late 1950s to the 1990s.[37] So began a new era of experimentation on GR and the EP.

In 1964 Roll, Krotkov and Dicke designed a torsion balance EP experiment which not only improved sensitivity but reduced noise as well.[38] Perhaps the most critical innovation was automatic measurement using an optical lever with its output recorded on paper. This eliminated a major source of error, the gravitational (and vibrational) disturbance from the operator taking measurements. Almost as importantly, the source of the acceleration was not terrestrial: Eötvös had used the centrifugal field of the Earth's rotation to balance a component of fall towards the center, which required turning the balance through 180 degrees, a major disturbance and source of uncertainty. Roll *et al* eliminated this by using fall toward the Sun as the source acceleration, balanced by Earth's orbital centrifugal force. Further, the apparatus was designed with a symmetry that would not couple efficiently to gravity gradient disturbances. Other innovations included operation in a high vacuum and attention to temperature fluctuations. Roll *et al* expressed their results in terms of what has become known as the "Eötvös parameter" $\eta$, the difference between test-body accelerations relative to their mean common acceleration. They found that $|\eta| < 1 \times 10^{-9}$ for copper and lead chloride while $|\eta| < 3 \times 10^{-11}$ for gold and aluminum. Significantly, they noted: "An interesting experimental possibility would be to put an apparatus in an artificial satellite. In this case, the advantage of the large force anomaly could be combined with that of a long observation time."

These results inspired significant effort in new ideas and experimental variations. For example, in 1968 Kreuzer tested the difference between active and passive mass for fluorine and bromine using a torsion balance with a teflon source suspended in a bromoform mixture.[39] Braginskii in 1972 claimed to have surpassed Roll *et al*'s sensitivity with an optical/photographic measurement.[40,41] He further optimized the torsion balance and used a laser beam technique to record its oscillations on photographic paper on a drum.

Space travel entered the picture in 1971, when astronaut David Scott performed a dramatic if imprecise free-fall test by dropping a feather and a hammer on the Moon in front of a worldwide television audience. Reflectors left on the lunar surface by three Apollo crews and two Soviet landers later enabled extraordinarily precise measurements of the Earth-Moon distance.[42] These could be used to compare the accelerations of the Earth and Moon toward the Sun, allowing one to test not only the EEP (because the Earth has an iron-nickel core while the Moon is primarily composed of silicate)[43] but also the SEP (because a larger fraction of the mass of the Earth is attributable to gravitational self-energy, a phenomenon sometimes referred to as the Nordtvedt effect[44]). Lunar laser ranging now constrains violations of the EEP at the $2 \times 10^{-13}$ level, implying an upper limit of $3 \times 10^{-4}$ for the SEP.[45]

Innovative new proposals for earthbound tests appeared throughout the 1970s, like one using diffraction patterns to compare bound and free-falling neutrons[46] and another based on elementary particle spin.[47] Work has continued into the 1980s on more-or-less classical tests and other gravitational phenomena at the University of Washington[48,49] and elsewhere.[50] Torsion balances have constrained EP violations by beryllium, copper and aluminum falling toward

Galactic dark matter,[51] copper and lead falling toward three-ton blocks of uranium,[52] and quartz crystals of different chirality.[53] The strongest such limit is currently $|\eta| < 2 \times 10^{-13}$ for beryllium and titanium,[54] comparable to that from lunar ranging.

Beginning in the 1990s, there was significant progress in tests based on atom interferometry,[55] which may be limited only by fundamental properties of atomic transitions. These currently constrain EP violations at the $7 \times 10^{-12}$ level for different isotopes of rubidium,[56] and there are proposals to compare rubidium and ytterbium,[57] and even atoms in entangled quantum states.[58] Completely new kinds of terrestrial tests continue to appear, like one based on transitions between opposite-parity states of atomic dysprosium.[59,60] Most recently, it has become possible to answer an old question[61] by verifying that antimatter falls the same way as ordinary matter using antihydrogen atoms at CERN.[62]

**Space Tests**

The first formal proposal for an orbiting EP test was by Chapman and Hanson in 1970.[63] As already noted by Roll *et al*, such a test would gain enormously in signal strength relative to earthbound torsion balance tests by using the gravitational field of the Earth, rather than the thousand times smaller field of the Sun or the centrifugal force of the Earth's rotation. It would further benefit from the elimination of seismic disturbances.

Roll, Krotkov and Dicke's experiment was essentially a rotating torsion balance orbiting the Sun, with disturbances from the Earth supporting it. But the Earth's gravity gradient is quite large in Earth orbit. Gravity gradients due to the spacecraft enclosing the experiment are also large, and both are variable. These disturbances rule out an orbiting torsion balance experiment of any practical size. Chapman realized this and proposed a sensitive accelerometer at the center of a massive but weightless rotating wheel, sized and proportioned to avoid coupling to gravity gradients. He suggested performing this experiment in the payload bay of the space shuttle which was then in early development stages.

In exchanges with Francis Everitt and Paul Worden at Stanford it became clear that further work was needed to achieve the full potential of an orbiting EP experiment. After a brief collaboration with Chapman, they adapted his proposal into a more practical design, eliminating the wheel and returning conceptually to Galileo's free-fall test. This would produce another factor of a thousand gain relative to terrestrial drop tests because orbiting masses would effectively "fall" for many thousands of seconds. The experiment could be protected from unwanted forces in orbit by drag-free control, reducing noise correspondingly. In this technique, disturbances are canceled by proportional thrusters that minimize spacecraft motion relative to a test mass, giving an almost perfect geodesic trajectory.

This work was funded for study by NASA from 1972 to 2007 to develop the basic technology needed, along with an extensive set of requirements for maximum sensitivity.[64,65,66] The project became known as STEP (Satellite Test of the EP) in 1989. The instrument was a cryogenic accelerometer optimized to test the universality of free fall between several pairs of nested cylindrical test masses of differing composition and a sensitivity to differences in acceleration approaching one part in $10^{18}$.

The STEP concept was close to Galileo (Fig. 2): drop masses from a tower with height equivalent to the Earth's radius but throw them horizontally (after Newton!) so fast that they miss the ground and fall all the way around — i.e., an orbit. If the EP is violated, two appropriately chosen test bodies follow different orbits. As is usually the case with sensitive experiments, the ideal concept is not quite adequate to the task of building it, and compromises must be made. A weak spring constant, stabilized by superconductors, was added to control and measure the masses. The final design depended heavily on technology borrowed from Gravity Probe-B,[67] including-drag free control to reduce disturbances, cryogenics for superconducting shielding, ultrahigh vacuum, and Superconducting Quantum Interference Device (SQUID)-based position sensors, plus a unique design for a single-axis superconducting differential accelerometer insensitive to gravity gradients to second (and later third) order.[68] Copper, gold, lead, rare earths and hydrogen (either solid or in compounds with other light elements like lithium, carbon or oxygen) were initially considered as test materials in 1993.[69] In 2000, the issue had not yet been settled,[70] and by 2007 it had been decided to use four test-mass pairs consisting of three materials (beryllium, niobium and platinum-iridium).[71]

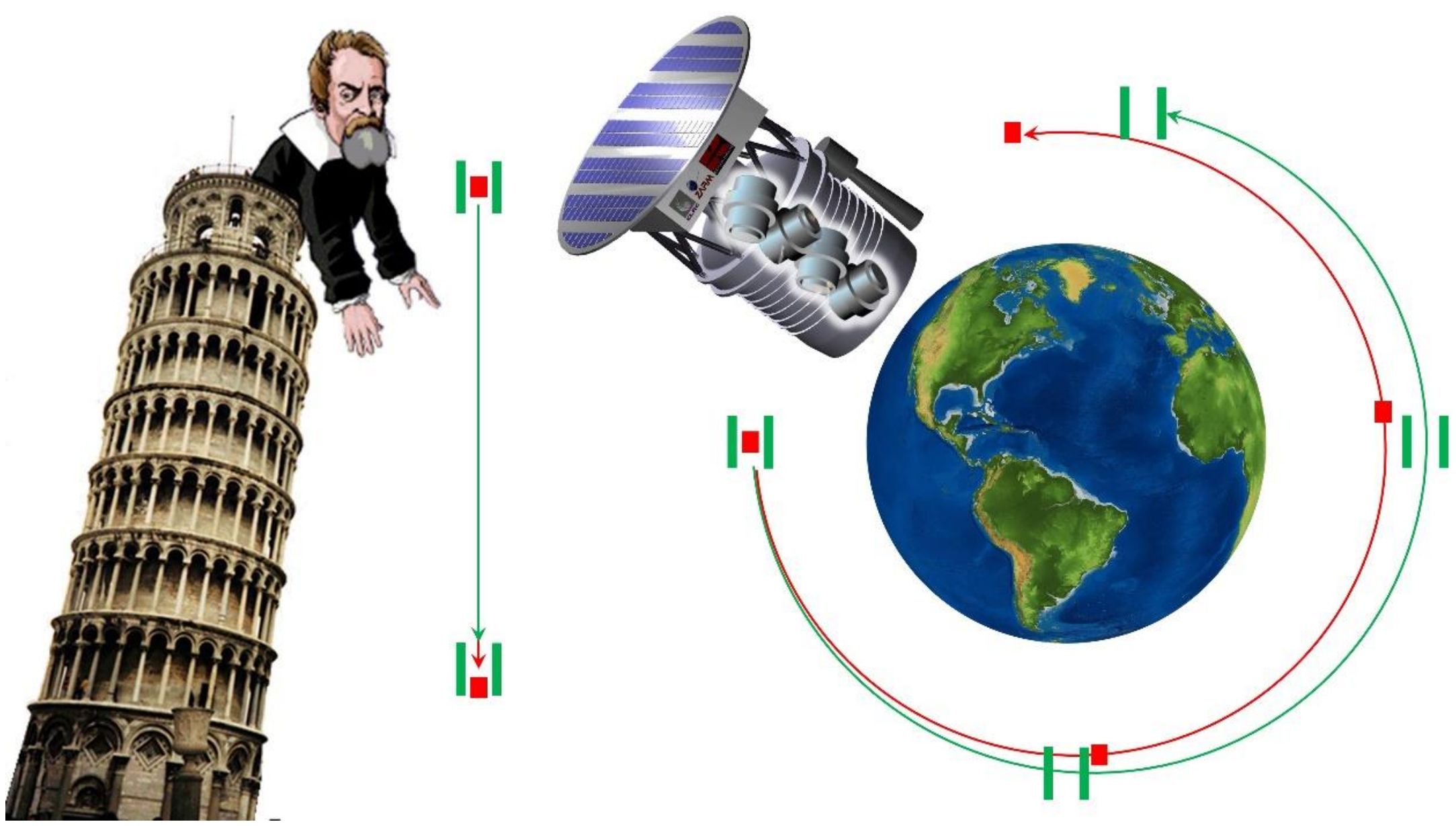

Fig. 2: STEP concept: an orbiting version of Galileo’s legendary drop test at Pisa

Interest in basic research waned among the American public and elected officials during the 1980s and 1990s, culminating in the cancellation of the Superconducting Super Collider in 1993. The basic requirements for a successful STEP mission had been demonstrated by late 1989, but NASA lost its appetite for fundamental physics experiments, especially ones that many expected might “only” produce a null result. The short-sightedness of such a view was amply proven by a century of revolutionary discoveries, from Michelson and Morley in 1887 to CP (Charge-Parity) violation in 1964; and it would soon be demonstrated again with the shocking detection of dark energy (i.e., a nonzero cosmological constant) in 1993. Nevertheless, STEP became a victim of this mindset, and moved to Europe.

In early 1990, STEP entered the European Space Agency (ESA)’s competition for M2-class missions, forming collaborations with teams in France, Germany, Great Britain, Italy, and the Netherlands. Notably, Pierre Touboul's group at ONERA (l’Office National d'Etudes et de Recherches Aérospatiales, the French space agency) was interested and joined the effort. STEP was unsuccessful but tried again for the next ESA medium-sized mission as a purely European venture. This effort in turn stimulated new ideas for gravitational tests in space. The Galileo Galilei experiment, also proposed as an M3-class mission, planned to use coupled cylinders in supercritical rotation perpendicular to their axis as an accelerometer,[72] while the Satellite Energy Exchange experiment proposed to use two small test bodies in nearby orbits.[73] These objects move in "horseshoe" orbits relative to one another, and their trajectories on closest approach give important information regarding Newton's gravitational constant $G$ as well as the EP. Seeds were sown at this time for what would eventually become the French/ESA project MICROSCOPE (MICROSatellite pour l'Observation du Principe d'Equivalence). The literature holds numerous other references to spin-off ideas from these beginnings.[74,75,76,77,78,79,80,81]

ONERA had its own accelerometer based on capacitive sensing from a previous gradiometer mission. During the competition for M2-class missions, Touboul hoped to use this in STEP.[82] As a non-cryogenic system, this instrument would not be sensitive to accelerations smaller than $10^{-15}$, especially after some necessary design changes. But it had the advantages of being simpler, already largely developed, and well supported. It was eventually used in MICROSCOPE, which launched in 2016 and took data from 2017-2018. Using test masses composed of platinum/rhodium and titanium/aluminum/vanadium alloys, MICROSCOPE detected no EP violations at the $2 \times 10^{-14}$ level, ten times better than existing terrestrial tests.[83,84,85] The importance of this mission in establishing the viability of space tests of the EP cannot be overstated. But their full potential remains to be realized.

## **Sensitivity and Robustness**

Nearly all experimental work on the EP to date has expressed results in terms of the dimensionless Eötvös parameter $\eta$, the relative difference in acceleration between test bodies in the same gravitational field. The focus on this quantity is natural and reflects the historical heritage of gravitational experiments with macroscopic masses. However, it is misleading from the perspective of modern particle physics and has the unfortunate effect of convincing some people that the EP has been so well tested that further progress is not needed. If $10^{-14}$ is already "close to zero," do we really need to go to $10^{-18}$?

The answer is yes! The reason lies in the microphysical origin of EP violations, as began to be appreciated in the 1990s by particle physicists working on unified theories of fundamental interactions (string theory) and cosmology with the newly discovered cosmological constant (dark energy). In all such theories, there are new fields that can in principle couple to gravity, leading to generic EP violations. But the relevant dimensionless quantity is not the macroscopic Eötvös parameter. It is the inherent *coupling strength* $(\alpha)$ of the new fields to those of the standard model, especially the gluons of Quantum Chromodynamics (QCD).[86,87] (This is logical, because most of what we measure as "mass" is actually nuclear binding energy.) The Eötvös parameter is tiny because it goes as the *square* of this underlying coupling parameter (just as the amplitude of a Feynman diagram goes as the square of the vertex factor representing the interaction between two fields). In the case of a single, generic new EP-violating scalar field, a coupling strength $\alpha \sim 10^{-6}$ implies an Eötvös parameter of order $10^{-13}$, while $\alpha \sim 10^{-8}$ implies $\eta \sim 10^{-18}$.[88] So the question we should be asking is: given that no EP-violating fields have been detected that couple to the standard model at the $10^{-6}$ level, do we need to go to $10^{-8}$?

To answer this question, we need look no farther than the strong CP problem within particle physics. As our governing theory of strong nuclear interactions, QCD is a cornerstone of the standard model. It contains a dimensionless free parameter, known as the CP-violating angle $\theta$, that is perfectly analogous to the EP-violating coupling strength $\alpha$. This parameter can take any value in principle, because CP-violating processes are not forbidden by QCD theory (any more than they are by electroweak theory). Yet CP violation occurs in weak interactions, but not in strong ones. Why not? Following an adage attributed to Murray Gell-Mann, "in physics, whatever is not forbidden must occur," experimentalists have invested great effort in searching for CP violation in strong interactions. Measurements of the electric dipole moment of the neutron now push upper limits on $\theta$ down to the $10^{-10}$ level.[89] At this level, particle theorists have begun to conclude that QCD theory must be incomplete, and that some new process must act to suppress CP violation dynamically. Such a process can be shown to be equivalent to the introduction of a new particle in nature, the *axion*, which has now become a leading dark matter candidate and the object of detection efforts around the world.

There is no essential difference between CP violation in strong interactions and EP violation in gravitational ones. Of course, GR is based on the EP, but at some level, we expect that GR must be reconciled with the rest of physics (the standard model of particle physics). Which will give? Can standard-model fields too be "geometrized" like the gravitational one? Or is it GR that must be extended? A breakdown of the EP would signal a new fundamental force of nature, a finding every bit as momentous as CP violation. Yet it has not received the same attention. In our view, the reason for that probably lies in the gulf that continues to divide the gravitational, astronomical and particle-physics communities, both theoretically and experimentally. This is a gulf that must be better bridged.

But when it comes to EP tests, sensitivity is only half of the story. The other, and possibly even more important half is robustness. We do not know what form EP violations will take. We have compared test masses composed of beryllium, aluminum and titanium with torsion balances, the Earth and Moon with lunar laser ranging, platinum and titanium in space, different isotopes of rubidium using atom interferometry, and matter and antimatter at CERN. But what if we really need to be comparing quarks and neutrinos, or bosons and fermions, or dark matter and dark energy? Given the range of possibilities, our only course is to design EP experiments to be not only as sensitive as possible, but also to use the widest possible range of test materials.

In practice, experimentalists are limited by considerations such as machinability, durability, and cost. A comprehensive effort to confront this issue was undertaken by Damour and Blaser beginning in the 1990s.[90,91,92] Motivated primarily by string theory, they focused on three kinds of "elementary charge" that might plausibly act as potential sources for EP violation while also discriminating usefully between the widest possible range of workable test materials. These are baryon number, neutron excess, and electromagnetic self-energy. Subsequent study has identified strong nuclear binding energy as another promising candidate.[93] Fig. 3 plots many common elements and compounds in the phase space defined by the first three of these properties. Platinum and titanium, the test materials used in MICROSCOPE, occupy almost the same spot in this diagram. STEP was designed to compare both elements with beryllium, spanning a significantly greater volume in phase space.

There was significant discussion within the STEP team about including a fourth material such as silicon. Four different test-mass pairs could have maximized EP-violating potential. But robustness also means being able to confirm or rule out systematic errors. A good way to accomplish this is to use a cyclic condition: given three accelerometers comparing materials A,B,C, the sum of differential acceleration measurements (A-B)+(B-C)+(C-A) should be zero.
A conservative approach is to use the fourth accelerometer as a control, measuring the same material A-A' rather than a fourth material. That way, if a violation is discovered, and the cyclic sum and control are both zero, one can be more certain that it is real.

However it is done, it is imperative for experimentalists to design and carry out EP tests spanning the greatest possible volume in potential EP-violating phase space, and for theorists to continue to extend this space in new directions. This applies not only to "active" experiments involving torsion balances and orbiting accelerometers, but also to "passive" tests using observations of the motions of solar-system bodies. Fig. 4 shows the same phase space as Fig. 3, but now populated by the rocky inner planets, gas giants and icy moons of the solar system, as well as the asteroid Psyche, which is thought to be largely composed of iron and is scheduled to be visited by a dedicated NASA mission. We recommend that EP-violating parameters be incorporated into the solar-system ephemeris that is maintained in the U.S.A. by the Jet Propulsion Laboratory (JPL). That ephemeris, which now relies mainly on radio tracking of spacecraft, is accurate to the centimeter level and has been successfully used to constrain the values of the PPN parameters $\beta$ and $\gamma$.[94] Some preliminary work has been done extending these results to EP violation based on "posterior fits" to existing ephemeris uncertainties,[95] but rigorous constraints can only be obtained from a self-consistent fit to all the data that incorporates the new parameters from the outset. Such constraints could be helpful in guiding experimentalists in their choice of test materials.

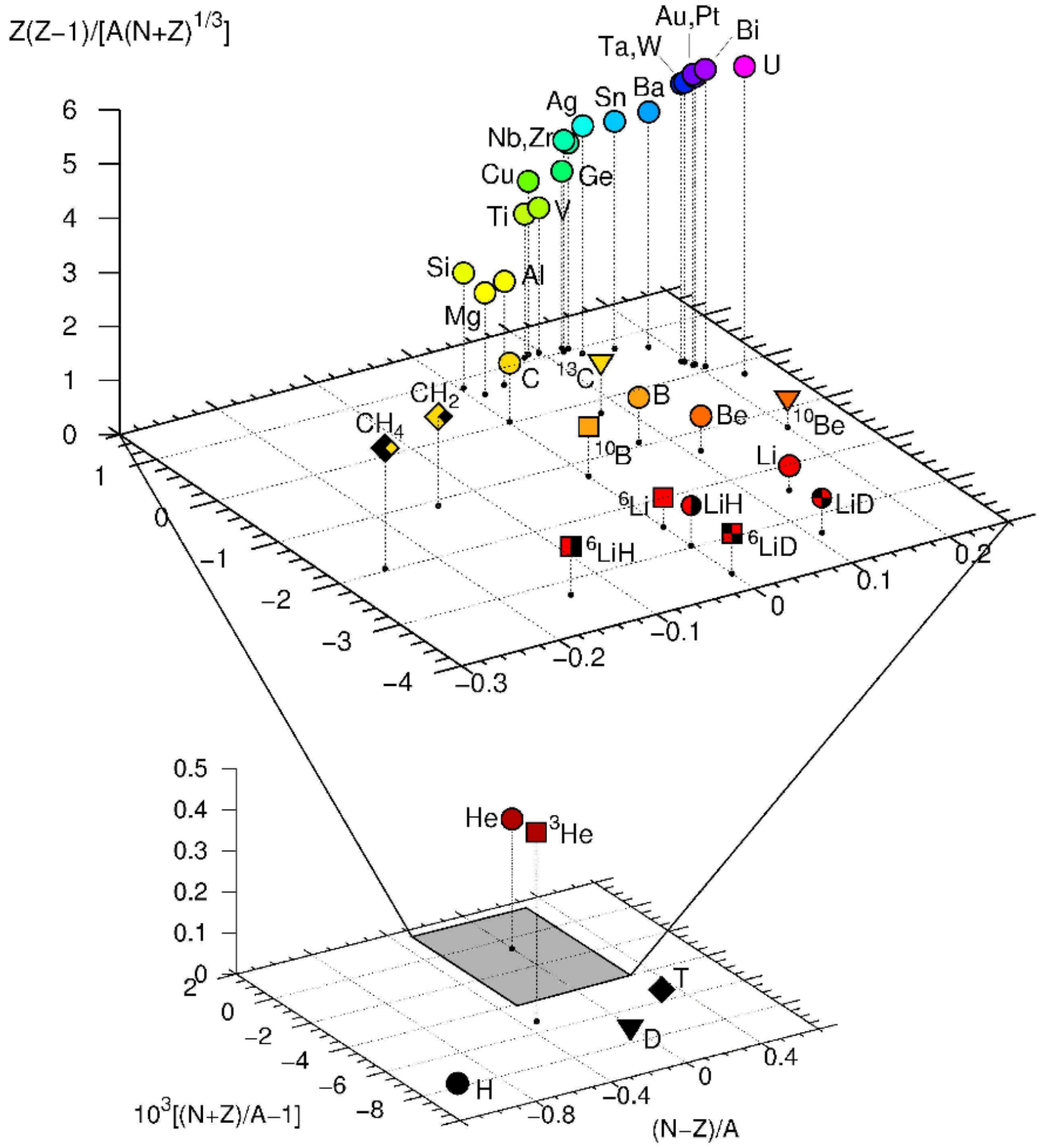

Fig. 3: Common elements and compounds plotted in the potential EP-violating phase space defined by baryon number $N+Z$, neutron excess $N-Z$ and nuclear electrostatic energy $[\propto Z(Z-1)]$, normalized by atomic mass $A$.[88]

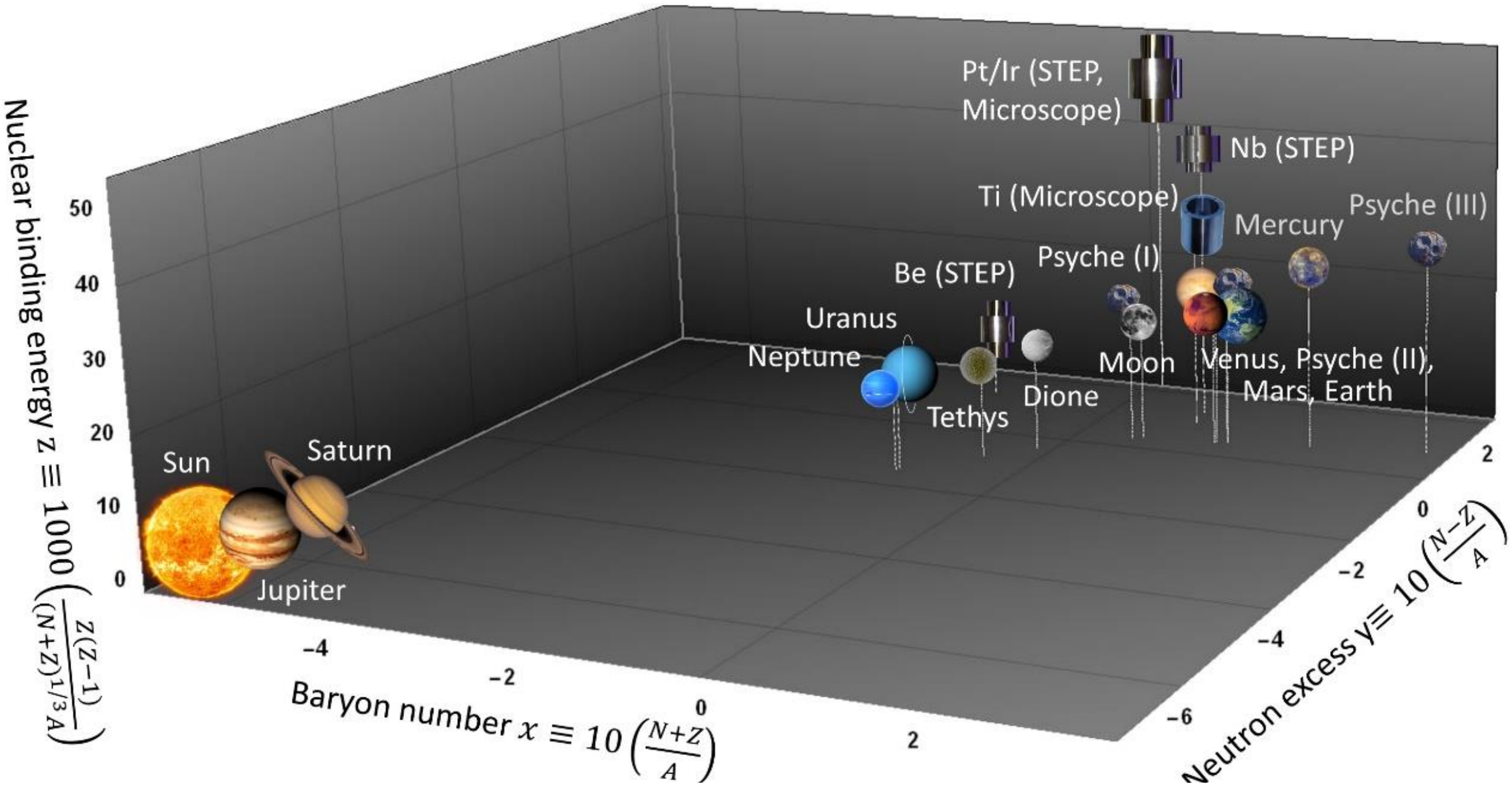

Fig. 4: Similar to Fig. 3 but now plotting a selection of solar-system bodies and space test materials based on composition.

**Observations on the process**

The brief history related above illustrates several broad trends:

- First: progress in ideas (theories) motivates progress in measurements, and vice versa.
- Second: progress in measurements follows progress in instrumentation, which may be based on new theory but is not necessarily driven by it.
- Third: progress comes in bursts or steps, which follow periods with little change.
- Fourth: the rate of change is accelerating.
- Fifth: connectivity is a significant factor in all of the preceding.

Progress in theories and progress in measurements — tests of those theories — naturally go together in science. A theory which describes nature inadequately cannot stand in the face of contradicting data, and will be strained by data that it doesn't fit, motivating a better theory. So progress in theories or ideas follows better measurements, because the measurements may discover new facts that were undreamed of in older philosophies. Then interest in new tests will develop from the new ideas, completing the cycle.

There is a smaller secondary cycle: better measurements naturally follow improvements in instrumentation and vice versa; but those improvements don't usually come from the pure desire to make better measurements to test a theory. There is another factor equally or more important, which we might call the socioeconomic motive. This is what drives and enables the inventors/innovators to make the effort in the first place. All the major contributors to instrumentation had both personal and social motivations, and adequate funds for their work, whether from the work itself, private fortune, or various sponsors.

The history of EP measurements with pendulums is an example of economics driving instrument development. Timekeeping, needed for navigation, led to development of better pendulums, which was further driven by measurements of the Earth's gravity needed to calibrate pendulum clocks, as well as interest in the shape of the Earth. This mainly economic motive drove the development of accurate pendulum clocks until the early 20th century, after which they suddenly became obsolete. The incidental improvement in pendulum EP measurements was enabled by the need for accurate time, but did not drive it.

The ability to measure accurate time without reference to the stars has become crucial in many other scientific measurements, although pendulums are no longer part of the process. At present it is hard to say whether science or economics is the major driver in timekeeping; Automatic navigation — presently at the decimeter level or better — critically depends on it, but scientific research interest is equally intense, in part because of its application to astronomical-scale measurements on gravitational waves and other phenomena.

In contrast, the next improvement in EP testing came from the development of torsion balances for gravity gradiometry. The torsion balance originated in Coulomb's instrument for determining the force between charges, and Cavendish's instrument used to "weigh the Earth" was derived from it. Torsion balances had (and still have) relatively little practical or economic value outside of gravity gradiometry. Major advances came less from commercial development than from scientific need, especially in the late 20th century.

The third observation: progress comes in bursts. This to be expected when a new idea for either instrumentation or theory appears. In the case of instruments, the idea usually results in significantly improved performance or the sudden ability or motivation to make a new type of measurement. Examples of improved performance are seen in the application of the torsion balance to EP; and much later, atom interferometers allow a completely new type of comparison, involving the fall of individual atoms to very high precision. "Breakthroughs" occur in both theoretical and experimental contexts when new concepts and methods appear.

The rate of change is accelerating, due largely to improvements in communication and the numbers of people involved. Months or even years could be needed for an exchange of letters and books in antiquity. By Galileo's time letters could be exchanged in weeks or less, and for physical delivery this decreased to a limit of a few days by the mid-19th century. For most of the 20th century colleagues and competitors were only minutes away by telephone. In consequence, the cycles of theory and experiment have accelerated: roughly 2000 years from Aristotle to Newtonian mechanics, another 200 to General Relativity, 20 more to quantum mechanics. The speedup has become limited by the rate at which information can be absorbed, evaluated, and worked into a new experiment, rather than the time to transmit it.

The final observation is that social connectivity is an essential part of these cycles. The picture is badly blurred by selection effects. The most prominent names in history are associated with great deeds and discoveries. These are naturally emphasized, even if only by the perpetrators, and further exaggerated by political and other interests. Hence the true origins of ideas and inventions often become concealed, lost or inaccessible. Still, ideas can often be traced between the prominent names and their contemporaries.

Connectivity was important in classical philosophy, as evidenced by philosophers arguing for and against each others' positions in their writings. But from antiquity even major publications may be lost or distorted, and letters between contemporaries in that era are generally not preserved, if they existed. It is almost as hard to document contemporary connections several hundred years ago or more. We can infer the existence of a broad network of researchers with mutual interest, in contact through publications, letters and word of mouth, from (e.g.) the spread of news of Galileo's astronomical discoveries. Ideas spread slowly by modern standards, but spread they certainly did. Galileo could very well have gotten his interest in the law of free fall from a slightly older contemporary, Simon Stevin, who studied motion, weights and falling bodies and did a drop experiment in 1586. There was very likely a reciprocal connection from Galileo and his father Vincenzo to Stevin, through work on musical semitones.[96] Stevin originated several physical and mathematical ideas including friction, an explanation of the tides, and the resolution of forces, but he is eclipsed by his better-known contemporary. A similar situation occurred around Newton;

many of the ideas that he assembled into one picture were already in circulation among his colleagues and contemporaries.

In modern instances, where papers and correspondence are more easily available, connectivity can often be uncovered quickly. Again, the EP history provides examples. Development of modern orbital EP tests from Dicke's suggestion resulted in multiple proposed space missions. The investigators were connected in multiple ways including correspondence, telephone, publications and scientific meetings, the effects of which were further multiplied by the speed of modern communication.

These trends are all well illustrated by the history of the EP. Confused and incomplete ideas regarding speed, acceleration, resistance and force prevailed initially; Aristotle brought some order to them, but he misunderstood motion badly enough that his description of falling disagreed with common observation. Simple drop experiments allowed a new idea, impetus, to prevail; and further, better experiments led to a better understanding of accelerated motion and inertia, and ultimately to a very good understanding of the laws of nature. The new mechanics of Galileo, Newton and others enabled better technology, including clocks, balances, and telescopes. This in turn enabled better measurements.

The EP was the key to Einstein's thinking when he developed GR. It was already latent in Newton's physical laws. But like Aristotle fixating on the impossibility of a vacuum, Newton was stuck with concepts of absolute space and time, even as he extended gravity's influence to the Moon. Others in his circle and earlier had imagined a similar leap, but without an understanding of motion. Newton put the two together and they pretty nearly fit. His imagination was limited to absolutes, but his achievement was breakthrough enough.

Einstein did something similar. Having put in the work of developing SR, he was conceptually positioned to extend its scope in a way that most of his contemporaries were not. He had "only" to see the connection with free fall and learn how to express the results in a form that did not depend on coordinates, even for accelerating observers. It fit very well indeed. New experiments, new exchanges of ideas, and new "happiest moments" will break the logjam currently besetting fundamental physics and allow the cycle to begin all over again.

## ACKNOWLEDGEMENTS

The authors' views have been shaped by all those who met with us and/or visited or participated in the Theory Group associated with the Gravity Probe-B and STEP experiments at Stanford University over the years, including Ron Adler, Sean Carroll, Thibault Damour, Francis Everitt, David Gross, Leopold Halpern, Alan Kostelecký, John Mester, Holger Müller, Ken Nordtvedt, Alex Silbergleit, and many others.

---

[4] Carlo Rovelli: (2015). "Aristotle's physics: a physicist's look," *Journal of the American Philosophical Association 1* (2015), 23–40.

[5] Robert J. Whitaker, "Aristotle is not dead: Student understanding of trajectory motion," *American Journal of Physics 51*, No. 4 (1983), 352-357.

[6] Lucretius, Titus Lucretius Carus (trans. Ronald Edward Latham), *On the Nature of the Universe*, (London: Penguin, 1994), Book II, lines 235 *ff*.

[7] Stephen Greenblatt, *The Swerve: How the World Became Modern* (New York: W.W. Norton & Company, 2011).

[8] M.R. Cohen and I.E. Drabkin, *A Source Book in Greek Science* (Harvard University Press, 1948), p. 220.

[9] M.R. Cohen and I.E. Drabkin, *ibid*, p. 223.

[10] James MacLachlan, "Galileo's experiments with pendulums: real and imaginary," *Annals of Science 33* (1976) 173-185.

[11] Stillman Drake, *Galileo at Work: His Scientific Biography* (University of Chicago Press, 1978), pp. 20 *ff*.

[12] Galileo Galilei (trans. H. Crew and A. deSalvio), *Dialogues Concerning Two New Sciences* (Dover Books, 1954) pp. 153 *ff*.

[13] Isaac Newton and Florian Cajori, *Sir Isaac Newton's Mathematical Principles of Natural Philosophy and His System of the World* (Berkeley: University of California Press, 1934).

[14] Allan Franklin and Ronald Laymon, *Measuring Nothing, Repeatedly* (San Rafael: Morgan & Claypool, 2019), Chapter 3.

[15] W.L. Harper, S.R. Valluri and R.B. Mann, "Jupiter's moons as a test of the equivalence principle," in V. Gurzadyan et al (eds.), proc. Ninth Marcel Grossmann Meeting (Singapore: World Scientific, 2002) 1803-1813.

[16] C.W.F. Everitt, T. Damour, K. Nordtvedt and R. Reinhard, "Historical perspective on testing the Equivalence Principle," *Advances in Space Research 32*, No. 7 (2003) 1297-1300.

[17] Friedrich Wilhelm Bessel, "Versuche über die Kraft, mit welcher die Erde Körper von verschiedener Beschaffenheit anzieht," *Astronomische Nachrichten 10* (1832), 97.

[18] Loránd Eötvös, "Untersuchungen über Gravitation und Erdmagnetismus," *Annalen der Physik* (1896), 354-400.

[19] Damian Kreichgauer, "Einige Versuche über die Schwere," *Verhandlungen der physikalischen Gesellschaft zu Berlin 10* (1891) 13-16.

[20] J.H. Poynting and P.L. Gray, "An experiment in search of a directive action of one quartz crystal on another," *Philosophical Transactions of the Royal Society of London A192* (1899), pp. 245-256.

[21] L. Southerns, "Experimental investigation as to dependence of gravity on temperature," *Proceedings of the Royal Society of London. Series A78*, No. 525 (1906), 392-403.

[22] Leonard Southerns, "A determination of the ratio of mass to weight for a radioactive substance," *Proceedings of the Royal Society of London. Series A84*, No. 571 (1910), 325-344.

[23] Albert Einstein, "Document 31: Ideas and Methods. II. The theory of General Relativity," in Michel Janssen et al (eds.), *The Collected Papers of Albert Einstein. Volume 7: The Berlin Years: Writings, 1918–1921* (Princeton University Press, 2002).

[24] L.I. Schiff, "On experimental tests of the general theory of relativity," *American Journal of Physics 28* (1960) 340-343.

[25] Alan P. Lightman and David L. Lee, "Restricted proof that the weak equivalence principle

implies the Einstein equivalence principle," *Physical Review D8*, No. 2 (1973), 364.

[26] Clifford M. Will, The confrontation between General Relativity and experiment," *Living Reviews in Relativity 17* (2014).

[27] James Overduin, Francis Everitt, Paul Worden and John Mester, "The science case for STEP," *Advances in Space Research 43* (2009) 1532.

[28] R.F.C. Vessot, M.W. Levine *et al*, "Test of relativistic gravitation with a space-borne hydrogen maser," *Physical Review Letters 45* (1980) 2081.

[29] Kenneth Nordtvedt, Jr., "Equivalence principle for massive bodies. II. Theory," *Physical Review 169*, No. 5 (1968), 1017.

[30] Charles W. Misner, Kip S. Thorne and John Archibald Wheeler, *Gravitation* (Macmillan, 1973).

[31] V.A. Kostelecký and J. Tasson, "Prospects for large relativity violations in matter-gravity couplings," *Physical Review Letters 102* (2009) 010402.

[32] Jay D. Tasson, "The Standard-Model Extension and gravitational tests," *Symmetry 8* (2016) 111.

[33] Lijing Shao and Quentin G. Bailey, "Testing the gravitational weak equivalence principle in the standard model extension with binary pulsars," *Physical Review D99* (2019) 084017.

[34] Pieter Zeeman, "Some experiments on gravitation. The ratio of mass to weight for crystals and radioactive substances," *Koninklijke Nederlandse Akademie van Wetenschappen Proceedings B20* (1918), 542-553.

[35] Loránd Eötvös, Desiderius Pekár and Eugen Fekete, "Beiträge zum Gesetze der Proportionalität von Trägheit und Gravität," *Annalen der Physik 373*, No. 9 (1922), 11-66.

[36] Harold H. Potter, "Some experiments on the proportionality of mass and weight," *Proceedings of the Royal Society of London A104,* No. 728 (1923), 588-610.

[37] Google Books Ngram Viewer, http://books.google.com/ngrams

[38] Peter G. Roll, Robert Krotkov and Robert H. Dicke, "The equivalence of inertial and passive gravitational mass," *Annals of Physics 26*, No. 3 (1964), 442-517.

[39] I.B. Kreuzer, "Experimental measurement of the equivalence of active and passive gravitational mass," *Phys. Rev. 199* (1968), 1007-12.

[40] Vladimir B. Braginski and Vladimir I. Panov, "The equivalence of inertial and gravitational masses," *Sov. Phys. Usp. 14* (1972), 811.

[41] Vladimir Borisovich Braginskii and Anatoliĭ Borisovich Manukin, *Measurement of Weak Forces in Physics Experiments*, (University of Chicago Press, 1977).

[42] C.O. Alley, R.F. Chang, D.G. Curri, J. Mullendore, S.K. Poultney, J.D. Rayner, E.C. Silverberg, C.A. Steggerda, H.H. Plotkin, W. Williams, B. Warner, H. Richardson and B. Bopp, "Apollo 11 laser ranging retro-reflector: initial measurements from the McDonald Observatory," *Science, Vol. 167*, No. 3917 (23 January 1970), pp. 368-370.

[43] J.G. Williams, R.H. Dicke, P.L. Bender, C.O. Alley, W.E. Carter, D.G. Currie, D.H. Eckhardt *et al*, "New test of the equivalence principle from lunar laser ranging," *Physical Review Letters 36*, No. 11 (1976), 551.

[44] Kenneth Nordtvedt, "Lunar laser ranging and laboratory Eötvös-type experiments," *Physical Review D37* (1988) 1070.

[45] Jürgen Müller, Thomas W. Murphy, Ulrich Schreiber, Peter J. Shelus, Jean-Marie Torre, James G. Williams, Dale H. Boggs, Sebastien Bouquillon, Adrien Bourgoin and Franz Hofmann, "Lunar Laser Ranging: a tool for general relativity, lunar geophysics and Earth

science," *Journal of Geodesy 93*, No. 11 (2019), 2195-2210.

[46] I. Koester, "Verification of the equivalence of gravitational and inertial mass for the neutron," *Phys. Rev. D14* (1976), 907-9.

[47] Asher Peres, "Test of equivalence principle for particles with spin," *Physical Review D18,* No. 8 (1978), 2739.

[48] Eric G. Adelberger, "Constraints on composition-dependent interactions from the Eöt-Wash experiment," in O. Fackler and J. Trân Thanh Vân (eds.), *5th Force Neutrino Physics,* proc. $8^{th}$ Moriond Workshop (Gif-sur-Yvette: Editions Frontières, 1988), pp. 485-499.

[49] C.W. Stubbs, "Testing the equivalence principle in the field of the Earth: an update on the Eöt-Wash experiment," in O. Fackler and J. Trân Thanh Vân (eds.), *New and Exotic phenomena,* proc. $10^{th}$ Moriond Workshop (Gif-sur-Yvette: Editions Frontières, 1990): 225-232.

[50] J.E. Faller, T.M. Niebauer and M.P. McHugh, "Current research efforts to test the equivalence principle at short ranges," in O. Fackler and J. Trân Thanh Vân (eds.), *5th Force Neutrino Physics,* proc. $8^{th}$ Moriond Workshop (Gif-sur-Yvette: Editions Frontières, 1988), p. 457.

[51] G. Smith, E. G. Adelberger, B. R. Heckel, and Y. Su, "Test of the equivalence principle for ordinary matter falling toward dark matter," Physical Review Letters 70 (1993) 123.

[52] J.H. Gundlach, G.L. Smith, E.G. Adelberger, B.R. Heckel, and H.E. Swanson, "Short-range test of the Equivalence Principle," *Physical Review Letters 78* (1997) 2523.

[53] L. Zhu, Q. Liu, H.-H. Zhao, Q.-L. Gong, S.-Q. Yang, P. Luo, C.-G. Shao, Q.-L. Wang, L.-C. Tu and J. Luo, "Test of the Equivalence Principle with chiral masses using a rotating torsion pendulum," *Physical Review Letters 121* (2018) 261101.

[54] S. Schlamminger, K.-Y. Choi, T. A. Wagner, J. H. Gundlach, and E. G. Adelberger, "Test of the Equivalence Principle using a rotating torsion balance," *Physical Review Letters 100* (2008) 041101.

[55] Mark Kasevich and Steven Chu, "Measurement of the gravitational acceleration of an atom with a light-pulse atom interferometer," *Applied Physics B54*, No. 5 (1992), 321-332.

[56] Peter Asenbaum, Chris Overstreet, Minjeong Kim, Joseph Curti and Mark A. Kasevich, "Atom-interferometric test of the equivalence principle at the $10^{-12}$ level," *Physical Review Letters 125*, No. 19 (2020), 191101.

[57] J. Hartwig, S. Abend, C. Schubert, D Schlippert, H. Ahlers, K. Posso-Trujillo, N. Gaaloul, W. Ertmer and E.M. Rasel, "Testing the universality of free fall with rubidium and ytterbium in a very large baseline atom interferometer," *New Journal of Physics 17* (2015) 035011

[58] Remi Geiger and Michael Trupke, "Proposal for a quantum test of the weak equivalence principle with entangled atomic species," *Physical Review Letters 120* (2018) 0043602.

[59] Arman Cingoz, "A laboratory test of Einstein's Equivalence Principle in atomic dysprosium: search for temporal and gravitational variation of the fine-structure constant" (Ph.D. thesis, University of California, Berkeley), *Dissertation Abstracts International,* Vol. 70-10, Section B, p. 6281 (2009).

[60] M.A. Hohensee, N. Leefer, D. Budker, C. Harabati, V.A. Dzuba and V.V. Flambaum, "Limits on violations of Lorentz symmetry and the Einstein Equivalence Principle using radio-frequency spectroscopy of atomic dysprosium," *Physical Review Letters 111* (2013) 050401.

[61] F.C. Witteborn and W.M. Fairbank, "Experimental comparison of the gravitational force on freely falling electrons and metallic electrons," *Physical Review Letters 19* (1967) 1049.

[62] E.K. Anderson *et al*, "Observation of the effect of gravity on the motion of antimatter," *Nature*

*621* (2023) 716-722.

[63] P.K. Chapman and A.J. Hanson, "An Eötvös experiment in Earth orbit," in R. Wayne Davies (ed.), *Proceedings of the Conference on Experimental Tests of Gravitation Theories*, November 11-13, 1970, California Institute of Technology, Pasadena, California. Jet Propulsion Laboratory Vol. 33, No. 499 (1970), 228.

[64] P.W. Worden Jr. and C.W.F. Everitt, "The gyroscope experiment. III: Tests of the equivalence of gravitational and inertial mass based on cryogenic techniques," in B. Bertotti (ed.), *Experimental Gravitation*, proc. International School of Physics "Enrico Fermi," Course XVI (Varenna, Italy, July 1972) (New York: Academic Press, 1974) 381-402.

[65] Paul W. Worden, Jr., "Equivalence principle tests in Earth orbit," *Acta Astronautica 5*, No. 1-2 (1978), 27-42.

[66] Paul W. Worden, Jr., "Almost exactly zero: the Equivalence Principle," in J.D. Fairbank, B.S. Deaver, Jr., C.W.F. Everitt and P.F. Michelson (eds.), *Near Zero: New Frontiers of Physics* (New York: W.H. Freeman and Company, 1988) 766-783.

[67] C.W.F. Everitt et al, "The Gravity Probe B test of General Relativity," *Classical and Quantum Gravity 32* (2015) 224001.

[68] N.A. Lockerbie and J.-P. Blaser, "Helium tides and test-mass design," in Rüdiger Reinhard (ed.), *STEP Symposium: Testing the Equivalence Principle in Space*, Proceedings of an International Symposium Held in Pisa, Italy, 6-8 April 1993 (ESA Publications Division, 1996), 225-238.

[69] Paul W. Worden Jr., "Satellite Test of the Equivalence Principle," in M. Demianski and C.W.F. Everitt (eds.), *The First William Fairbank Meeting on Relativistic Gravitational Experiments in Space* (Singapore: World Scientific, 1993) 43-59.

[70] P. Worden, R. Torii, J.C. Mester, and C.W.F. Everitt, "The STEP payload and experiment," *Advances in Space Research 25* (2000) 1205-1208.

[71] Paul W. Worden, on behalf of the STEP team, "STEP payload development," *Advances in Space Research 39* (2007) 259-267.

[72] A.M. Nobili, D. Bramanti, E. Polacco, G. Catastini, A. Anselmi, S. Portigliotti, A. Lenti and A. Severi, "The 'Galileo Galilei (GG) project: testing the Equivalence Principle in space and on Earth," *Advances in Space Research 25* (2000) 1231-1235.

[73] Alvin J. Sanders and W.E. Deeds, "Proposed new determination of the gravitational constant *G* and tests of Newtonian gravitation," *Physical Review D46*, No. 2 (1992) 489.

[74] W. Vodel, H. Koch, S. Nietzsche, J.V. Zameck Glyscinski, R. Neubert and H. Dittus, "Testing Einstein's equivalence principle at Bremen Drop Tower using LTS SQUID technique," *IEEE Transactions on Applied Superconductivity 11*, No. 1 (2001) 1379-1382.

[75] J.D. Anderson and J.G. Williams, "Long-range tests of the equivalence principle," *Classical and Quantum Gravity 18*, No. 13 (2001) 2447.

[76] R. Bonneville, "GEOSTEP: A gravitation experiment in Earth-orbiting satellite to test the Equivalence Principle," *Advances in Space Research 32*, No. 7 (2003) 1367-1372.

[77] Timothy J. Sumner, "The STEP and GAUGE missions," *Space Science Reviews 148* (2009) 475-487.

[78] Robert D. Reasenberg and James D. Phillips, "A weak equivalence principle test on a suborbital rocket," *Classical and Quantum Gravity 27*, No. 9 (2010) 095005.

[79] Feng-Tian Han, Qiu-Ping Wu, Ze-Bing Zhou and Yuan-Zhong Zhang, "Proposed space test of the new equivalence principle with rotating extended bodies," *Chinese Physics Letters 31*, No. 11 (2014) 110401.